\documentclass[
showpacs,preprintnumbers,amsmath,amssymb]{revtex4}
\usepackage{amsmath}
\usepackage{graphicx}

\def\be{\begin{equation}}
\def\ee{\end{equation}}
\def\bea{\begin{eqnarray}}
\def\eea{\end{eqnarray}}

\def\e{{\rm e}}
\begin{document}

\title{Stability of cosmological solutions in F(R) Ho\v{r}ava-Lifshitz gravity}
\author{Diego S\'{a}ez-G\'{o}mez}
\affiliation{Institut de Ciencies de l'Espai (ICE-CSIC/IEEC),
Campus UAB, Facultat de Ciencies, Torre C5-Par-2a pl, E-08193
Bellaterra (Barcelona), Spain}

\begin{abstract}
The present paper is devoted to the analysis of cosmological solutions and its stability in the frame of $F(R)$ Ho\v{r}ava-Lifshitz gravity. The perturbations around general spatially flat FLRW solutions are analyzed and it is shown that the stability of those solutions  depends on the type of theory, i.e. on the form of the action $F(R)$, as well as on the extra parameters contained in every Ho\v{r}ava-Lifshitz theory (due to the  breaking of Lorentz invariance). The (in)stability of cosmological solutions can provide a constraint of the models and it may give new observational predictions. A natural explanation of the end of inflation and radiation/matter phases can be provided by this class of theories. An explicit example of  $F(R)$ gravity is studied, and the transition between the different epochs of the Universe history is achieved.
\end{abstract}


\maketitle

\section{Introduction}

Since observational data suggests that the Universe expansion is accelerating, a large number of models have been proposed to explain this phenomenon. In the frame of General Relativity (GR), the accelerating expansion can not be explained unless new terms or fields  are considered. The main candidate for dark energy is the so-called $\Lambda$CDM model, which incorporates a cosmological constant and late-time acceleration can be achieved. Nevertheless,  $\Lambda$CDM model contains several unresolved problems, as the fine tuning problem, such that other different proposals have been seriously considered, as for example scalar  fields (quintessence, phantom,..) or modifications of GR as the so called $F(R)$ gravity (see Refs.~\cite{Reviews} for reviews on unification of inflation and dark energy in modified gravity). At the same time, it is also accepted that another epoch of accelerated expansion, known as inflation, occurred  during the early Universe. This suggests that both epochs may be unified under the same mechanism. In the frame of $F(R)$ gravity, the unification of both epochs is easily achieved and it gives a natural explanation in terms  purely of gravity (see Refs.~\cite{Reviews, FR2, FR4, FRdeSitter,Reconstruction}).  

Recently, a new theory of gravity that claims to be power-counting renormalizable has been suggested in Ref.~\cite{Horava}. This new theory, already known as Ho\v{r}ava-Lifshitz gravity, breaks Lorentz invariance, what makes the theory to be renormalizable, but it produces  consequently a lot of problems. However, it is conjectured that the Lorentz invariance is recovered in the  IR limit (see Ref.~\cite{Horava3}). Some aspects of cosmology has already been studied in the frame of this new theory (see Ref.~\cite{cosm}). Nevertheless, as in General Relativity, Ho\v{r}ava-Lifshitz gravity can not explain dark energy epoch as well as inflation without new terms or fields, remaining  such problem unresolved. An extension of the standard $F(R)$ gravity to Ho\v{r}ava-Lifshitz theory has been performed (see Refs.~\cite{FRhorava, FRhorava2, FRhorava3, kluson}), which seems to be also renormalizable, and  late-time acceleration can be reproduced without cosmological constant or any other exotic field  (see Refs.~\cite{FRhorava,FRhorava2}). Even  the unification of dark energy epoch and inflation can be performed in this new class of theories, and the so-called viable $F(R)$  models, which avoid violations of the local gravity tests, can be easily extended to Ho\v{r}ava-Lifshitz gravity (see Ref.~\cite{FRhorava3}). 

At the current paper, cosmological solutions of the type of spatially flat Friedmann-Lema\^{\i}tre-Robertson-Walker (FLRW)  are studied in the frame of F(R) Ho\v{r}ava-Lifshitz gravity, and their stability  is analyzed. In particular, we  focus on the study of stability of radiation/matter dominated eras, where the Universe expands by a power law, and de Sitter solutions, which can well  describe the accelerated expansion epochs of the Universe history. We explore  spatially independent perturbations around these solutions, where the effects of the extra terms incorporated  by the function $F(R)$ are studied. Also the new parameters included in the theory, due to the breaking of the Lorentz invariance, could affect the cosmological solutions. The (in)stability of  a solution gives very important information, as the possible exit from a phase of the cosmological history, constraints on the kind of action $F(R)$ and/or future observational predictions. An explicit example of a $F(R)$ action, where an unstable de Sitter solution is found, is studied. This model performs a successful exit from inflation, and produces an instability at the end of matter dominated epoch, such that a phase transition may occur.  

The paper is organized as follows: in the next section, $F(R)$ Ho\v{r}ava-Lifshitz gravity is briefly introduced, and the cosmological equations are obtained. Sect. III is devoted to the analysis of general spatially flat FLRW solutions and their stability.  The general equation for the perturbations in the linear approach is obtained. In particular, the class of solutions described by a scale factor that depends on a power of time (radiation/matter dominated epochs) are analyzed. In Appendix A, de Sitter solutions are also studied  in detail. In Sect. IV an explicit example of $F(R)$ gravity is studied, where one of the so-called viable models, that  unifies dark energy and inflationary epochs, is analyzed. Finally, some discussions and conclusions are provided in the last section.

\section{Framework}
In this section, modified Ho\v{r}ava-Lifshitz $F(R)$
gravity is briefly reviewed (see Refs.~\cite{FRhorava, FRhorava2, FRhorava3, kluson}). We start by writing a
general metric in the so-called ADM decomposition in a $3+1$ spacetime (for
more details see Refs.~\cite{ADM},\cite{gravitation} and references
therein),
\be
ds^2=-N^2 dt^2+g^{(3)}_{ij}(dx^i+N^idt)(dx^j+N^jdt)\, ,
\label{1.1}
\ee
where $i,j=1,2,3$, $N$ is the so-called lapse variable, and $N^i$ is
the shift $3$-vector. In standard General Relativity,
the Ricci scalar can be written in terms of this metric, and yields
\be
R=K_{ij}K^{ij}-K^2+R^{(3)}+2\nabla_{\mu}(n^{\mu}\nabla_{\nu}n^{\nu}-n^{\nu}
\nabla_{\nu}n^{\mu})\, ,
\label{1.2}
\ee
here $K=g^{ij}K_{ij}$, $K_{ij}$ is the extrinsic curvature, $R^{(3)}$
is the spatial scalar curvature, and $n^{\mu}$ a unit vector
perpendicular to a hypersurface of constant time. The extrinsic
curvature $K_{ij}$ is defined as
\be
K_{ij}=\frac{1}{2N}\left(\dot{g}_{ij}^{(3)}-\nabla_i^{(3)}N_j-\nabla_j^{(3)}
N_i\right)\, .
\label{1.3}
\ee

In the original Ho\v{r}ava-Lifshitz model \cite{Horava}, the lapse variable $N$ is taken
to be just time-dependent, so that the projectability condition holds
and by using the foliation-preserving diffeomorphisms (\ref{1.7}),
it can be fixed to be $N=1$.
As it is pointed out in \cite{Blas:2009qj},  imposing the projectability
condition may cause problems with Newton's law in Ho\v{r}ava gravity.
On the other hand, Hamiltonian analysis shows that the non-projectable
$F(R)$-model is inconsistent (see Ref.~\cite{masud}).
For the non-projectable case, the Newton law
could be restored (while keeping stability) by the ``healthy''
extension of the original Ho\v{r}ava gravity of Ref.~\cite{Blas:2009qj}.

The action for standard $F(R)$ gravity can be written as
\be
S=\int d^4x\sqrt{g^{(3)}}N F(R)\, .
\label{1.4}
\ee
Ho\v{r}ava-Lifshitz gravity is assumed to have different scaling
properties of the space and time coordinates
\be
x^i=b x^i\, , \quad t=b^zt\, ,
\label{1.6}
\ee
where $z$ is a dynamical critical exponent that renders the theory
renormalizable for $z=3$ in $3+1$ spacetime dimensions as it is shown in \cite{Horava}
(for a proposal of covariant renormalizable gravity with dynamical
Lorentz symmetry breaking, see \cite{no}).
GR is recovered when $z=1$. The scaling properties (\ref{1.6}) render
the theory  invariant only under the so-called foliation-preserving 
diffeomorphisms:
\be
\delta x^i=\zeta(x^i,t)\, , \quad \delta t=f(t)\, .
\label{1.7}
\ee
It has been pointed out that, in the IR limit, full diffeomorphisms
are recovered, although the mechanism for this transition is not
physically clear. The action considered here was introduced
in Ref.~\cite{FRhorava},
\be
S=\frac{1}{2\kappa^2}\int dtd^3x\sqrt{g^{(3)}}N F(\tilde{R})\, , \quad
\tilde{R}= K_{ij}K^{ij}-\lambda K^2 + R^{(3)}+
2\mu\nabla_{\mu}(n^{\mu}\nabla_{\nu}n^{\nu}-n^{\nu}\nabla_{\nu}n^{\mu})-
L^{(3)}(g_{ij}^{(3)})\, ,
\label{1.8}
\ee
where $\kappa$ is the dimensionless gravitational coupling, and where, two
new constants $\lambda$ and $\mu$ appear, which account for the violation
of full diffeomorphism transformations.
A degenerate version of the above $F(R)$-theory with $\mu=0$ has been proposed and
studied in Ref.~\cite{kluson}.
Note that in the original Ho\v{r}ava gravity theory \cite{Horava},
the third term in the expression for $\tilde{R}$ can be omitted, as
it turns out to be a total derivative. The term $L^{(3)}(g_{ij}^{(3)})$ is
chosen to be \cite{Horava}
\be
L^{(3)}(g_{ij}^{(3)})=E^{ij}G_{ijkl}E^{kl}\, ,
\label{1.9}
\ee
where the generalized De Witt metric is given by,
\be
G^{ijkl}=\frac{1}{2}\left(g^{(3)ik}g^{(3)jl}+g^{(3)il}g^{(3)jk}\right)-\lambda g^{(3)ij}g^{(3)kl}\, .
\label{1.10}
\ee
In Ref.~\cite{Horava}, the expression for $E_{ij}$ is constructed to
satisfy the ``detailed balance principle'' in order to restrict the
number of free parameters of the theory, and it is defined through the
variation of an action
\be
\sqrt{g^{(3)}}E^{ij}=\frac{\delta W[g_{kl}]}{\delta g_{ij}}\, ,
\label{1.11}
\ee
where the form of $W[g_{kl}]$ is given in Ref.~\cite{Horava2} for
$z=2$ and  $z=3$. Other forms for
$L^{(3)}(g_{ij}^{(3)})$ have been suggested that abandons the
detailed balance condition but still render the theory power-counting
renormalizable (see Ref.~\cite{FRhorava2}).

We are interested in the study of cosmological
solutions for the theory described by action (\ref{1.8}).
Spatially-flat FLRW metric is assumed
\be
ds^2=-N^2dt^2+a^2(t)\sum_{i=1}^3 \left(dx^{i}\right)^2\, .
\label{1.14}
\ee
If we also assume the projectability condition,
$N$ can be taken to be just time-dependent and, by using the
foliation-preserving
diffeomorphisms (\ref{1.7}), it can be set to  unity, $N=1$.
When we do not assume the projectability condition, $N$ depends on both 
the time and spatial coordinates. Then, just as an assumption of
the solution, $N$ is taken to be unity.

For the metric (\ref{1.14}), the scalar $\tilde{R}$ is
given by
\be
\tilde{R}=\frac{3(1-3\lambda
+6\mu)H^2}{N^2}+\frac{6\mu}{N}\frac{d}{dt}\left(\frac{H}{N}\right)\, .
\label{1.15}
\ee
For the action (\ref{1.8}), and assuming the FLRW metric (\ref{1.15}),
the second FLRW equation can be obtained by varying
the action with respect to the spatial metric $g_{ij}^{(3)}$, what
yields
\be
0=F(\tilde{R})-2(1-3\lambda+3\mu)\left(\dot{H}+3H^2\right)F'(\tilde{R})-2(1-
3\lambda)H
\dot{\tilde{R}}F''(\tilde{R})+2\mu\left(\dot{\tilde{R}}^2F^{(3)}(\tilde{R})
+\ddot{\tilde{R}}F''(\tilde{R})\right)+\kappa^2p_m\, ,
\label{1.16}
\ee
here $\kappa^2=16\pi G$, $p_m$ is the pressure of a perfect fluid
that fills the Universe, and $N=1$. Note that this
equation turns out the usual second FLRW equation for standard $F(R)$
gravity (\ref{1.4}) when  $\lambda=\mu=1$.
If we assume the projectability condition,
variation over $N$ of the action (\ref{1.8}) yields the following
global constraint
\be
0=\int d^3x\left[F(\tilde{R})-6F'(\tilde{R})\left\lbrace(1-3\lambda +3\mu)H^2+\mu\dot{H}\right\rbrace +6\mu
H \dot{\tilde{R}}F''(\tilde{R})-\kappa^2\rho_m\right]\, .
\label{1.17}
\ee
Now,  by using the ordinary conservation equation for the matter fluid
$\dot{\rho}_m+3H(\rho_m+p_m)=0$, and integrating Eq.~(\ref{1.16}), it yields
\be
0=F(\tilde{R})-6\left[(1-3\lambda
+3\mu)H^2+\mu\dot{H}\right]F'(\tilde{R})+6\mu H
\dot{\tilde{R}}F''(\tilde{R})-\kappa^2\rho_m-\frac{C}{a^3}\, ,
\label{1.18}
\ee
where $C$ is an integration constant, taken to be zero,
according to the constraint equation (\ref{1.17}).
If we do not assume the projectability condition, we can directly
obtain (\ref{1.18}), which corresponds to the first FLRW equation, by 
varying the action (\ref{1.8}) over $N$.
Hence, starting from a given $F(\tilde{R})$ function, and solving 
Eqs.~(\ref{1.16}) and (\ref{1.17}), a cosmological solution
can be obtained.

\section{Cosmological solutions and its stability in $F(\tilde{R})$ gravity}

In this section, we are interested to study the stability of general cosmological solutions in the frame of $F(\tilde{R})$ Ho\v{r}ava-Lifshitz gravity, with special attention to those cosmological solutions that shape the history of the Universe, as de Sitter or power law solutions. It is well known that in standard F(R) gravity, any cosmological solution can be reproduced by reconstructing the function  $F(R)$ (see Ref.~\cite{Reconstruction}). As it was shown in Ref.~\cite{FRhorava3}, dark energy and even the unification with the inflationary epoch can be reconstructed in this new frame of $F(\tilde{R})$ Ho\v{r}ava-Lifshitz theories. The stability of those solutions plays a crucial role in order to get the transition from one cosmological phase to another.

\subsection{Stability of general flat FLRW cosmological solutions}

Let us start by studying  a general spatially flat FLRW metric (\ref{1.14}).  We  focus specially on de Sitter and power law solutions of the type $a(t)\propto t^{m}$ because dark energy and radiation/matter dominated eras are governed by this class of cosmological solutions respectively. The implications of the extra geometrical terms coming from $F(\tilde{R})$ could be determinant for the stability and transition during the epochs of Universe evolution. Firstly, we assume a general solution, 
\be
H(t)=h(t) \ .
\label{2.7a}
\ee
Then, the scalar curvature $\tilde{R}$ yields,
\be
\tilde{R}_h(t)=  3(1-3\lambda+6\mu)h^2(t)+6\mu \dot{h}(t)\ .
\label{2.7}
\ee
Assuming  a certain $F(\tilde{R})$ that reproduces the solution (\ref{2.7a}),  the FLRW equation (\ref{1.18}) has to be fulfilled, 
\be
0=F(\tilde{R_h})-6\left[(1-3\lambda
+3\mu)h^2+\mu\dot{h}\right]F'(\tilde{R_h})+6\mu h
\dot{\tilde{R}}_hF''(\tilde{R_h})-\kappa^2\rho_m\, ,
\label{2.8}
\ee
where the matter fluid is assumed to be a perfect fluid with equation of state $p_m=w_m\rho_m$, where $w_m$ is a constant. By the energy conservation equation $\dot{\rho}_m+3h(1+w_m)\rho_m=0$, the evolution of the matter energy density can be expressed in terms of the solution $h(t)$ as,
\be
\rho_{mh}=\rho_0 \e^{-3(1+w_m)\int h(t)dt}\ ,
\label{2.9}
\ee
where $\rho_0$ is an integration constant. We are interested to study the perturbations around the arbitrary solution $h(t)$. For that purpose, let us expand the function $F(\tilde{R})$ in powers of $\tilde{R}$  around (\ref{2.7}),
\be
 F(\tilde{R})=F_h+F'_h(\tilde{R}-\tilde{R}_h)+\frac{F''_h}{2}(\tilde{R}-\tilde{R}_h)^2+\frac{F^{(3)}_h}{6}(\tilde{R}-\tilde{R}_h)^3+O(\tilde{R}-\tilde{R}_h)^4\ ,
\label{2.10}
\ee
where the derivatives of the function $F(\tilde{R})$ are evaluated at $R_h$, given in (\ref{2.7}). Note that matter perturbations also contribute to the stability, inducing a mode on the perturbation. Then, we can write the perturbed solution  as,
\be
H(t)=h(t)+\delta(t)\ , \quad \rho_m\simeq\rho_{mh}(1+\delta_m(t))\ .
\label{2.11}
\ee
Hence, by introducing the above quantities in the FLRW equation,  the equation for the perturbation $\delta(t)$  becomes (in the linear approximation),
\be
\ddot{\delta}+b\dot{\delta}+\omega^2\delta=\frac{\kappa^2\rho_{mh}}{36\mu^2hF_h''}\delta_m\ ,
\label{2.12}
\ee
where,
\[
 b=-\frac{h'}{h}-\frac{1-3\lambda+3\mu}{\mu}h+\frac{1-3\lambda+6\mu}{\mu}+6\left((1-3\lambda+6\mu)h\dot{h}+\mu\ddot{h}\right)\frac{F^{(3)}_h}{F_h''}\ ,
\]
\[
\omega^2=\left[1-3\lambda+6\mu-2(1-3\lambda+3\mu)h\right]\frac{F_h'}{6\mu^2hF_h''}
\]
\be
+(1-3\lambda+6\mu)\left(\frac{-1+3\lambda-3\mu}{\mu^2}h+\frac{-1+h}{\mu h}\dot{h}\right)+\frac{\ddot{h}}{h}+6(1-3\lambda+6\mu)\left(\frac{1-3\lambda+6\mu}{\mu}\dot{h}h+\ddot{h}\right)\frac{F^{(3)}_h}{F_h''}\ .
\label{2.12a}
\ee
In this case, the solution for $\delta(t)$ can be split in two branches, one corresponding to the homogeneous part of the equation (\ref{2.12}), whose solution will depend on the background theory, i.e. on $F(\tilde{R})$ and its derivatives, and another one corresponding to the particular solution of  eq.~(\ref{2.12}), which represents the term induced by  matter perturbation $\delta_m$. Then, the complete solution can be written as,
\be
\delta(t)=\delta_{homg}(t)+\delta_{inh}(t)\ .
\label{2.13}
\ee
We are interested in the perturbations induced  by the function $F(\tilde{R})$ and its derivatives, so that we focus on the homogeneous solution $\delta_{homg}$. By a first qualitative analysis, we can see that the homogeneous part of the equation (\ref{2.12}) yields exponential or damped oscillating perturbations.  The form of the perturbations  depends completely on the form of the function $F(\tilde{R})$ and its derivatives evaluated at $R_h$. Note that in general, the equation (\ref{2.12a}) has to be solved by numerical methods. Nevertheless, we  could assume some restrictions to obtain qualitative information. Let us consider the cases, 
\begin{itemize}

 \item The trivial case,  given by  $F_h'=F_h''=F_h^{(3)}=0$,  makes the perturbation tends to zero, $\delta(t)=0$, and the cosmological solution $h(t)$ is stable. 

\item For $F_h'\neq0$ and $F_h'',F_h^{(3)}\rightarrow0$, the term that dominates in (\ref{2.12}) is given by,
\be
\omega^2\sim\left[1-3\lambda+6\mu-2(1-3\lambda+3\mu)h\right]\frac{F_h'}{6\mu^2hF_h''}\ .
\label{2.13b}
\ee
And the stability of the cosmological solution $h(t)$ depends on the sign of this term, and therefore, on the model $F(\tilde{R})$  and the solution $h(t)$.

\item For $F_h',F_h''\rightarrow0$ but $F_h^{(3)}\neq0$, looking at (\ref{2.12a}),  the perturbation depends on the value of the last term in the coefficients $b$ and $\omega^2$, which can be approximated to,
\be
b\sim6\left((1-3\lambda+6\mu)h\dot{h}+\mu\ddot{h}\right)\frac{F^{(3)}_h}{F_h''}\ ,\quad \omega^2\sim 6(1-3\lambda+6\mu)\left(\frac{1-3\lambda+6\mu}{\mu}\dot{h}h+\ddot{h}\right)\frac{F^{(3)}_h}{F_h''}\ .
\label{2.13a}
\ee
 The cosmological solution will be stable in the case that both  coefficients (\ref{2.13a}) are greater than zero, what yields a damped oscillating perturbation that decays.

\end{itemize}
However, in general the  equation (\ref{2.12}) can not be solved  analytically for arbitrary solutions $h(t)$ and actions $F(\tilde{R})$, and numerical analysis is required.  Nevertheless, by imposing certain conditions on $F(\tilde{R})$ as above,  qualitative information can be obtained. In order to perform a deeper analysis, some specific solutions $h(t)$ are studied below, as well as an explicit example of $F(\tilde{R})$.

\subsection{Stability of radiation/matter eras: Power law solutions}

In this section,  an important class of cosmological solutions is considered, the power law solutions, which are described by the Hubble parameter,
\be
H(t)=\frac{m}{t} \quad \rightarrow \quad a(t)\propto t^m\ .
\label{2.14}
\ee
In the context of General Relativity, this class of solutions are generated by a perfect fluid with equation of state parameter $w=-1+\frac{2}{3m}$, and the matter/radiation dominated epochs are approximately described by  (\ref{2.14}). Also phantom epochs can be described by this class of solutions when $m<0$ . Let us study the stability for the Hubble parameter (\ref{2.14}), and how the inclusion of extra terms in the action and the new parameters ($\lambda, \mu$) may affect the stability of the solution (\ref{2.14}).  As in the above section,  the perturbation equation (\ref{2.12}) can not  be solved analytically in general, although under some restrictions  we can obtain important qualitative information about the stability of the solution. Then, by assuming an $F(\tilde{R})$ that approximately does not deviate from Hilbert-Einstein action during radiation/matter dominated epochs, the  second and third derivatives can  be neglected $F''_h,F^{(3)}_h\sim0$  (as they must become important only during  dark energy epoch and/or inflation). In such a case, the coefficient in front of $\delta(t)$ in the eq.~(\ref{2.12}) is approximated as,
\be
\omega^2\sim\left[1-3\lambda+6\mu-2(1-3\lambda+3\mu)h(t)\right]\frac{F_h'}{6\mu^2h(t)F_h''}\ .
\label{2.15}
\ee
Then, the value of the frequency $\omega^2$ depends on the time, such that the stability may change along the phase. For small values of $t$, the frequency takes the form $\omega^2\sim-2(1-3\lambda+3\mu)\frac{6\mu^2F_h'}{F_h''}$,  and  assuming $\lambda\sim\mu$, the perturbations will grow exponentially when   $\frac{F_h'}{F_h''}>0$, and  the solution becomes unstable.  While for large $t$, the frequency can be approximated as $\omega^2\sim(1-3\lambda+6\mu)\frac{F_h'}{6\mu^2h(t)F_h''}$ and the instability will be large if $\frac{F_h'}{F_h''}<0$, and  a phase transition may occur.

\section{Example of a viable $F(\tilde{R})$ model}

Let us consider an explicit model of  $F(\tilde{R})$ gravity in order to apply the analysis about the stability performed above. We are interested to study the stability of radiation/matter dominated eras as well as de Sitter solutions for an explicit $F(\tilde{R})$. Here we consider a model proposed in Ref.~\cite{FR2}, and studied in Ref.~\cite{FR4} in the context of standard gravity and generalized to Ho\v{r}ava-Lifshitz gravity in Ref.~\cite{FRhorava3}. The action is defined as,
\be
F(\tilde{R})=\chi\tilde{R}+\frac{\tilde{R}^n(\alpha\tilde{R}^n-\beta)}{1+\gamma\tilde{R}^n}\, ,
\label{3.1}
\ee 
where ($\chi, \alpha, \beta, \gamma$) is a set of constant parameters of the theory. In standard gravity, this model can reproduce well late-time acceleration with no need of a cosmological constant or any kind of exotic field, as well as also  inflation, such that the unification of both epochs of the Universe history under the same mechanism can be performed (see Ref.~\cite{FR2}). For simplicity,  we assume $n=2$ in (\ref{3.1}) for our analysis. The radiation/matter dominated epochs, which can be described by the class of solutions given in (\ref{2.14}),  could suffer a phase transition to the era of dark energy due to the instabilities caused by the second term of the action (\ref {3.1}). Then, we are interested to study the possible effects produced by the presence of these extra geometric terms during the cosmological evolution. By assuming the solution (\ref{2.14}), and following the steps described in the above section, the stability is affected by the derivatives of the function (\ref{3.1}) evaluated in $h(t)=m/t$. We are interesting in large times, when the end of matter dominated epoch has to occur. At that moment the derivatives of $F(\tilde{R})$ can be approximated as,
\be
F_h'\rightarrow \chi\ , \quad F_h''\rightarrow -2\beta\ , \quad F_h^{(3)}\rightarrow0\ .
\label{3.2}
\ee 
Here for simplicity, we have assumed  $0<\beta<<1$. By means of the analysis performed in the previous section, we can conclude that the linear perturbation $\delta(t)$  grows exponentially, and the radiation/matter dominated phase becomes unstable for large times, what may produce the transition to another different phase. Then, the  $F(\tilde{R})$ function (\ref{3.1}) can explain perfectly the end of matter dominated epoch with no need of the presence of a cosmological constant. \\

Let us now study the stability of de Sitter solutions  (for more details on de Sitter solutions and its stability, see Appendix A). It is known that the  model (\ref{3.1}) may contain several de Sitter solutions (see Ref.~\cite{FRhorava3} and \cite{FR4}), solutions of the first FLRW equation that now turns out an algebraic equation given by,
\be
\tilde{R}_0+\frac{\tilde{R}_0^n(\alpha\tilde{R}_0^n-\beta)}{1
+\gamma\tilde{R}_0^n}
+\frac{6H_0^2(-1+3\lambda-3\mu)\left[1+n\alpha\gamma\tilde{R}_0^{3n-1}
+\tilde{R}_0^{n-1}(2\gamma\tilde{R}_0-n\beta)+\tilde{R}_0^{2n-1}(\gamma^2\tilde{R}_0
+2n\alpha)\right]}{(1+\gamma\tilde{R}_0^n)^2}=0\, .
\label{3.3}
\ee 
This  equation has to be solved numerically, even for the simple case studied here, $n=2$. Nevertheless, one of the de Sitter points from the model (\ref{3.1}) is defined by  a minimum of the second term in the action (\ref{3.1}). By assuming the constraint on the parameters $\beta\gamma/\alpha\gg 1$,  the minimum that represents a de Sitter point is given by,
\be
\tilde{R}_0 \sim \left( \frac{\beta}{\alpha\gamma}\right)^{1/4}\, ,
\qquad F'(\tilde{R}_0)=\chi\, , \qquad  F(\tilde{R}_0)= \tilde{R}_0-2\Lambda\ , \quad \text{where} \quad \Lambda\sim
\frac{\beta}{2\gamma}\, .
\label{3.5}
\ee
Then, by evaluating the derivatives of (\ref{3.1}) around $\tilde{R}_0$ and by the equation (\ref{2.6}), the perturbation $\delta(t)$ can be calculated. Note that the stability condition for  de Sitter solution,  given by $\frac{F_0'}{F_0''}>12H_0^2$ in Appendix A, is not satisfied for this case as $F_0''>>F_0'$, such that the de Sitter point (\ref{3.5}) is  unstable. By resolving eq.~(\ref{2.6}), the perturbation is given by exponential functions,
\be
\delta(t)=C_1\e^{a_+t}+C_2\e^{a_-t}\ , \quad \text{with} \quad a_{\pm}=\frac{H_0(1-3\lambda+3\mu)}{2\mu}\ .
\label{3.6}
\ee
Hence, the model (\ref{3.1}) is unstable around this de Sitter point (\ref{3.5}), what  predicts the exit from an accelerated phase in the near future, providing a natural explanation about the end of the inflationary epoch, or a future prediction about the end of dark energy era. However,  the theory described by (\ref{3.1}) may contain more de Sitter points, given by the roots of equation (\ref{3.3}), which may be stable. Then, a deeper analysis has to be performed to study the entire Universe evolution for this model of $F(\tilde{R})$ gravity.

\section{Discussions}
At the present paper, we have analyzed spatially flat FLRW cosmology for  nonlinear Ho\v{r}ava-Lifshitz gravity. Basically we have  extended  standard $F(R)$ gravity to Ho\v{r}ava-Lifshitz theory, which reduces to the first one in the IR limit (where we assume that the parameters $(\lambda,\mu)$  are reduced to unity). The stability of this general class of solutions has been studied and it is shown that it depends mainly on the choice of the function $F(\tilde{R})$ and in part on the values of the parameters $(\lambda,\mu)$. For large times, when the scalar curvature  is very small,  the main effect of the perturbation on a cosmological solution is caused by the value of the derivatives of $F(\tilde{R})$. It is shown that in general, the perturbation equation can not be solved analytically, even in the linear approach. Nevertheless, under some restrictions, important information is obtained, and the (in)stability of the different phases of the Universe history can be studied. For specific values of the derivatives of $F(\tilde{R})$, a given solution can becomes (un)stable, which means a major constraint on models. By analyzing an explicit example in Sect. IV, where an $F(\tilde{R})$ function of the class of viable models is considered, we have found that this kind of theories can well explain the end of matter dominated epoch, and reproduces late-time acceleration. We have shown that for this specific example, there is a de Sitter point that becomes unstable, what  predicts the end of this de Sitter epoch, providing a  natural explanation of the end of inflationary era. However, as this model (and in general every $F(\tilde{R})$ model) may contain several de Sitter solutions, where some of them can be stable, a further analysis of the phase space  has to be performed to connect the different regions of the Universe history. \\
Hence, the analysis made here provides a general approach for the study of spatially flat FLRW solutions in the frame of higher order Ho\v{r}ava-Lifshitz gravities, which can restrict the class of functions $F(\tilde{R})$ allowed by the observations, and it gives a natural explanation of the end of inflation and matter dominated epoch, shaping the Universe history in a natural way.

\begin{acknowledgments}

I would like to thank Emilio Elizalde and Sergei Odintsov for giving me support to perform this task and useful discussions. I acknowledge an FPI fellowship from MICINN (Spain), project FIS2006-02842. 

\end{acknowledgments}

\appendix

\section{de Sitter solutions in $F(\tilde{R})$  gravity}

Let us consider one of the simplest but most important solutions in cosmology, de Sitter (dS) solution. As dark energy and  inflation can be shaped (in its simplest form) by a dS solution, its stability becomes very important, specially in the case of inflation, where a successful exit is needed to end  the accelerated phase occurred during the  early Universe. In general,  standard $F(R)$ gravity contains several de Sitter points, which represent critical points (see \cite{FRdeSitter}). The analysis can be extended to  $F(\tilde{R})$ Ho\v{r}ava-Lifshitz gravity, where the de Sitter solution $H(t)=H_0$ has to satisfy the first equation FLRW equation (\ref{1.18}),
\be
0=F(\tilde{R}_0)-6H^2_0(1-3\lambda+3\mu)F'(\tilde{R}_0)\, ,
\label{2.1}
\ee
where we have taken $C=0$ and  assumed absence of any kind of matter. For this case, the scalar $\tilde{R}$ is given  by,
\be
\tilde{R}_0=3(1-3\lambda+6\mu)H_0^2\ .
\label{2.2}
\ee
Then, the positive roots of equation (\ref{2.1}) are de Sitter points allowed by a particular choice of a $F(\tilde{R})$ function. By assuming a de Sitter solution, we expand $F(\tilde{R})$  as a series of powers of the scalar $\tilde{R}$ around $\tilde{R}_0$, 
\be
 F(\tilde{R})=F_0+F'_0(\tilde{R}-\tilde{R}_0)+\frac{F''_0}{2}(\tilde{R}-\tilde{R}_0)^2+\frac{F^{(3)}_0}{6}(\tilde{R}-\tilde{R}_0)^3+O(\tilde{R}^4)\ .
\label{2.3}
\ee
Here, the primes denote derivatives respect $\tilde{R}$, while the subscript $0$ means that the function $F(\tilde{R})$ and its derivatives are evaluated at $R=\tilde{R}_0$. Then, by perturbing the solution,  the Hubble parameter can be writing as,
\be
H(t)=H_0+\delta(t)\ .
\label{2.4}
\ee
Using  (\ref{2.3}), and the perturbed solution (\ref{2.4}) in the first FLRW equation (\ref{1.18}), the equation for the perturbation yields,
\[
0=\frac{1}{2}F_0-3H_0^2(1-3\lambda +3\mu)-3H_0\left[\left((1-3\lambda)F_0'+6F_0''H_0^2(-1+3\lambda-6\mu)(-1+3\lambda-3\mu)\right)\delta(t) \right.
\]
\be
\left.+6F_0''\mu H_0(-1+3\lambda-3\mu)\dot{\delta}(t)-12F_0''\mu^2\ddot{\delta}(t)\right]\ .
\label{2.5}
\ee
Here we have restricted the analysis to the linear approximation on $\delta$ and its derivatives. Note that the first two terms in the equation (\ref{2.5}) can be removed because of equation (\ref{2.1}), which is assumed to be satisfied, and equation (\ref{2.5}) can be rewritten in a more convenient form as,
\be
\ddot{\delta}(t)+\frac{H_0(1-3\lambda+9\mu)}{2\mu}\dot{\delta}(t)+\frac{1}{12\mu^2}\left[(3\lambda-1)\frac{F_0'}{F_0''}-6H_0^2(1-3\lambda+6\mu)(1-3\lambda+3\mu)\right]\delta(t)=0\ .
\label{2.6}
\ee
Then, the perturbations on a  dS solution will depend completely on the model, specifically on the derivatives of  $F(\tilde{R})$, as well as on the parameters  $(\lambda,\mu)$. The instability becomes large if the term in front of $\delta(t)$ (the frequency) in the equation (\ref{2.6}) becomes negative and the perturbation grows exponentially, while if we have a positive frequency, the perturbation  behaves as a damped harmonic oscillator. During  dark energy epoch, when the scalar curvature is very small,  the IR limit of the theory can be assumed, where  $\lambda=\mu\sim1$, and the frequency  depends completely on the value of $\frac{F_0'}{F_0''}$. In order to avoid large instabilities during the dark energy phase, the condition  $\frac{F_0'}{F_0''}>12H_0^2$ has to be fulfilled. Nevertheless, when the scalar curvature is large, the IR limit is not a convenient approach, and the perturbation depends also on the values of $(\lambda,\mu)$. If we assume a very small $F_0''$,  the frequency in the equation (\ref{2.6})  dominates compared to the other terms, and by assuming $\lambda>1/3$, the stability of the solution will depend on the sign of $\frac{F_0'}{F_0''}$, being stable when it is positive.


\begin{thebibliography}{}
\bibitem{Reviews}
S.~Nojiri, S.~D.~Odintsov, arXiv:1011.0544  [gr-qc], 
arXiv:1008.4275 [hep-th], 
   eConf {\bf C0602061}, 06 (2006)
   [Int.\ J.\ Geom.\ Meth.\ Mod.\ Phys.\  {\bf 4}, 115 (2007)]
   [arXiv:hep-th/0601213]; \\
   S.~Capozziello, M.~De Laurentis and V.~Faraoni,
   arXiv:0909.4672 [gr-qc]. 
\bibitem{FR2}
   S.~Nojiri and S.~D.~Odintsov,
   Phys.\ Rev.\  D {\bf 77}, 026007 (2008)
   [arXiv:0710.1738 [hep-th]].
\bibitem{FR4}
   E.~Elizalde and D.~S\'aez-G\'omez,
   Phys.\ Rev.\  D {\bf 80}, 044030 (2009)
   [arXiv:0903.2732 [hep-th]].
\bibitem{FRdeSitter}G.~Cognola, E.~Elizalde, S.~D.~Odintsov, P.~Tretyakov, and S~Zerbini, Phys.\ Rev.\ D \textbf{79}, 044001 (2009) [arXiv:0810.4989 [gr-qc]].
\bibitem{Reconstruction}
Nojiri S., Odintsov S.D., S\'aez-G\'omez D., 
Phys. Lett. B \textbf{681} 74 (2009) [arxiv:0908.1269].
\bibitem{Horava}
   P.~Horava,
   Phys.\ Rev.\  D {\bf 79}, 084008 (2009)
   [arXiv:0901.3775 [hep-th]].
\bibitem{Horava3}
   P.~Horava, C.~M.~Melby-Thompson
  Phys.\ Rev.\ D {\bf 82}, 064027 (2010)
   [arXiv:1007.2410 [hep-th]].

\bibitem{cosm}
   T.~Takahashi and J.~Soda,
   Phys.\ Rev.\ Lett.\  {\bf 102}, 231301 (2009)
   [arXiv:0904.0554 [hep-th]]; \\
   E.~Kiritsis and G.~Kofinas,
   Nucl.\ Phys.\  B {\bf 821}, 467 (2009)
   [arXiv:0904.1334 [hep-th]]; \\
   R.~Brandenberger,
   Phys.\ Rev.\  D {\bf 80}, 043516 (2009)
   [arXiv:0904.2835 [hep-th]]; \\
   S.~Mukohyama, K.~Nakayama, F.~Takahashi and S.~Yokoyama,
   Phys.\ Lett.\  B {\bf 679}, 6 (2009)
   [arXiv:0905.0055 [hep-th]]; \\
   T.~P.~Sotiriou, M.~Visser and S.~Weinfurtner,
   JHEP {\bf 0910}, 033 (2009)
   [arXiv:0905.2798 [hep-th]]; \\
   E.~N.~Saridakis,
   Eur.\ Phys.\ J.\  C {\bf 67}, 229 (2010)
   [arXiv:0905.3532 [hep-th]]; \\
   M.~Minamitsuji,
   Phys.\ Lett.\  B {\bf 684}, 194 (2010)
   [arXiv:0905.3892 [astro-ph.CO]]; \\
   G.~Calcagni,
   Phys.\ Rev.\  D {\bf 81}, 044006 (2010)
   [arXiv:0905.3740 [hep-th]]; \\
   A.~Wang and Y.~Wu,
   JCAP {\bf 0907}, 012 (2009)
   [arXiv:0905.4117 [hep-th]]; \\
   M.~i.~Park,
   JHEP {\bf 0909}, 123 (2009)
   [arXiv:0905.4480 [hep-th]]; \\
   S.~Nojiri and S.~D.~Odintsov,
   Phys.\ Rev.\  D {\bf 81}, 043001 (2010)
   [arXiv:0905.4213 [hep-th]]; \\
   M.~Jamil, E.~N.~Saridakis and M.~R.~Setare,
   Phys.\ Lett.\  B {\bf 679}, 172 (2009)
   [arXiv:0906.2847 [hep-th]]; \\
   M.~i.~Park,
   JCAP {\bf 1001}, 001 (2010)
   [arXiv:0906.4275 [hep-th]]; \\
   C.~Bogdanos and E.~N.~Saridakis,
   Class.\ Quant.\ Grav.\  {\bf 27}, 075005 (2010)
   [arXiv:0907.1636 [hep-th]]; \\
   C.~G.~Boehmer and F.~S.~N.~Lobo,
   arXiv:0909.3986 [gr-qc]; \\
   I.~Bakas, F.~Bourliot, D.~Lust and M.~Petropoulos,
   Class.\ Quant.\ Grav.\  {\bf 27}, 045013 (2010)
   [arXiv:0911.2665 [hep-th]]; \\
   G.~Calcagni,
   JHEP {\bf 0909}, 112 (2009)
   [arXiv:0904.0829 [hep-th]]; \\
   S.~Carloni, E.~Elizalde and P.~J.~Silva,
   Class.\ Quant.\ Grav.\  {\bf 27}, 045004 (2010)
   [arXiv:0909.2219 [hep-th]]; \\
   X.~Gao, Y.~Wang, W.~Xue and R.~Brandenberger,
   JCAP {\bf 1002}, 020 (2010)
   [arXiv:0911.3196 [hep-th]]; \\
   Y.~S.~Myung, Y.~W.~Kim, W.~S.~Son and Y.~J.~Park,
   arXiv:0911.2525 [gr-qc]; \\
   E.~J.~Son and W.~Kim,
   arXiv:1003.3055 [hep-th]; \\
   A.~Wang,
   arXiv:1003.5152 [hep-th]; \\
   A.~Ali, S.~Dutta, E.~N.~Saridakis and A.~A.~Sen,
   arXiv:1004.2474 [astro-ph.CO];\\
 S.~Mukohyama, arXiv:1007.5199 [hep-th].
\bibitem{FRhorava}
   M.~Chaichian, S.~Nojiri, S.~D.~Odintsov, M.~Oksanen and A.~Tureanu, Class.\ Quantum\ Grav.\ {\bf 27}, 185021 (2010)
   [arXiv:1001.4102 [hep-th]]; 
\bibitem{FRhorava2}
   S.~Carloni, M.~Chaichian, S.~Nojiri, S.~D.~Odintsov, M.~Oksanen and
A.~Tureanu, Phys.\ Rev.\ D {\bf 82}, 065020 (2010)
   [arXiv:1003.3925 [hep-th]]; 
\bibitem{FRhorava3} E.~Elizalde, S.~Nojiri, S.~D.~Odintsov and D.~S\'aez-G\'omez, \textit{Unifying inflation with dark energy in modfied F(R) Ho\v{r}ava-Lifshitz gravity}, arXiv:1006.3387
\bibitem{kluson}
   J.~Kluson,
   Phys.\ Rev.\  D {\bf 81}, 064028 (2010)
   [arXiv:0910.5852 [hep-th]]; 

   J.~Kluson,
   arXiv:1002.4859 [hep-th].
\bibitem{ADM}
R. L. Arnowitt, S. Deser and C. W. Misner, arxiv:gr-qc/0405109;
C.~Gao,
    Phys.\ Lett.\  B {\bf 684}, 85 (2010)    [arXiv:0905.0310 [astro-ph.CO]].
%
 \bibitem{gravitation}
 C.~W.~Misner, K.~S.~Thorne, J.~A.~Wheeler,
 \textit{Gravitation}, W. H. Freeman and Company, 1973, San Francisco.


\bibitem{Blas:2009qj}
  D.~Blas, O.~Pujolas and S.~Sibiryakov,
  Phys.\ Rev.\ Lett.\  {\bf 104}, 181302 (2010)
  [arXiv:0909.3525 [hep-th]].

\bibitem{masud}
M.~Chaichian, M.~Oksanen, A.~Tureanu, 
arXiv:1006.3235 [hep-th].

\bibitem{no}
   S.~Nojiri and S.~D.~Odintsov, Phys.\ Lett.\ B\ {\bf 691}, 60 (2010)
   [arXiv:1004.3613 [hep-th]].

\bibitem{Horava2}
   P.~Horava,
   JHEP {\bf 0903}, 020 (2009)
   [arXiv:0812.4287 [hep-th]].



\end{thebibliography}
\end{document}